\pdfoutput=1
\documentclass[journal=jacsat,manuscript=article,layout=traditional]{achemso}
\setkeys{acs}{articletitle=true}

\usepackage{amsmath,amsfonts,amssymb} %
\usepackage{mathtools} %
\ifx\phone\undefined
\usepackage{wasysym} 
\else
\let\Phone\phone
\let\phone\relax
\usepackage{wasysym}

\let\phone\Phone
\let\Phone\relax
\fi
\usepackage{bm} %
\usepackage[bbgreekl]{mathbbol} %
\usepackage{graphicx} %
\usepackage[dvipsnames,table]{xcolor} %
\usepackage{tabularx} %
\usepackage[acronym]{glossaries} %
\usepackage{braket} %
\usepackage[binary-units=true,detect-weight=true]{siunitx} %
\usepackage{fancyhdr} %
\usepackage{microtype} %
\usepackage[caption=false]{subfig} %
\usepackage[byname]{smartref} %
\usepackage[breaklinks, %
colorlinks, linkcolor=Blue, citecolor=Blue, urlcolor=Blue]{hyperref} %
\usepackage[version=3]{mhchem} %
\usepackage{wrapfig} %
\AtBeginDocument{\usepackage{booktabs}} 

\DeclareSIUnit[number-unit-product = {\,}]\cal{cal}
\def\MyTitle{Estimating Phosphorescent Emission Energies in Ir(III)
  Complexes using Large-Scale Quantum Computing Simulations} %
\def\MyAuthora{Scott N. Genin} %
\def\MyAuthorb{Ilya G. Ryabinkin} %
\def\MyAuthorc{Nathan R. Paisley} %
\def\MyAuthord{Sarah O. Whelan} %
\def\MyAuthore{Michael G. Helander} %
\def\MyAuthorf{Zachary M. Hudson} %

\def\MySubject{} %

\hypersetup{%
  pdftitle={\MyTitle{}}, %
  pdfauthor={\MyAuthora{}, \MyAuthorb{}, \MyAuthorc{}, \MyAuthord{},
    \MyAuthore{} and \MyAuthorf{}}, %
  pdfsubject={\MySubject{}}, %
  pdfkeywords={phosphorescent materials, triplet-singlet gap
    simulations, density functional theory, quantum chemistry, quantum
    computing} %
}

\newcolumntype{Y}{>{\centering\arraybackslash}X}

\newacronym[longplural={degrees of freedom}, firstplural={degrees of
  freedom (DOF)}, plural={DOF}]{DOF}{DOF}{degree of freedom} %
\newacronym[longplural={equations of motion}, firstplural={equations
  of motion (EOM)}, plural={EOM}]{EOM}{EOM}{equation of motion} %

\newacronym{OLED}{OLED}{organic light-emitting diode}
\newacronym{NISQ}{NISQ}{noisy intermediate-scale quantum}

\newacronym{JW}{JW}{Jordan--Wigner} %
\newacronym{BK}{BK}{Bravyi--Kitaev} %

\newacronym{QPE}{QPE}{quantum phase estimation} %
\newacronym{VQE}{VQE}{variational quantum eigensolver} %
\newacronym{QMF}{QMF}{qubit mean-field} %
\newacronym{QCC}{QCC}{qubit coupled cluster} %
\newacronym{iQCC}{iQCC}{iterative qubit coupled cluster} %
\newacronym{PQA}{PQA}{parametrized quantum annealing} %
\newacronym{DIS}{DIS}{direct interaction set} %

\newacronym{CAS}{CAS}{complete active space} %
\newacronym{PES}{PES}{potential energy surface} %
\newacronym{PEC}{PEC}{potential energy curve} %
\newacronym{MO}{MO}{molecular orbital} %

\newacronym{CI}{CI}{configuration interaction} %
\newacronym{FCI}{FCI}{full configurational interaction} %
\newacronym{CASCI}{CASCI}{complete active space configurational
  interaction} %
\newacronym{MCSCF}{MCSCF}{multiconfigurational self-consistent
  field} %
\newacronym{CASSCF}{CASSCF}{complete active space self-consistent
  field} %
\newacronym{CC}{CC}{coupled cluster} %
\newacronym{UCC}{UCC}{unitary coupled cluster} %
\newacronym{UCCSD}{UCCSD}{unitary coupled cluster singles and
  doubles} %
\newacronym{CCSD}{CCSD}{coupled-cluster singles and doubles} %
\newacronym{CCSD-T}{CCSD(T)}{coupled-cluster singles and doubles and
  non-iterative triples} %
\newacronym{RHF}{RHF}{restricted Hartree--Fock} %
\newacronym{CIS}{CIS}{configuration-interaction singles} %
\newacronym{ROHF}{ROHF}{restricted open-shell Hartree--Fock} %
\newacronym{UHF}{UHF}{unrestricted Hartree--Fock} %
\newacronym{DMRG}{DMRG}{density-matrix renormalization group} %
\newacronym{DFT}{DFT}{density-functional theory} %
\newacronym{TDDFT}{TDDFT}{time-dependent density-functional theory} %
\newacronym{ENPT}{ENPT}{Epstein-Nesbet perturbation theory} %
\newacronym{MP}{MP}{M{\o}ller--Plesset perturbation theory} %
\newacronym{MP2}{MP2}{second-order M{\o}ller--Plesset perturbation theory} %
\newacronym{MRMP2}{MRMP2}{second-order multi-reference M{\o}ller--Plesset perturbation theory} %

\newacronym{SQP}{SQP}{sequential quadratic programming} %
\newacronym{MMA}{MMA}{method of moving asymptotes} %

\def\be{\begin{equation}} %
\def\ee{\end{equation}} %
\def\bea{\begin{eqnarray}} %
\def\eea{\end{eqnarray}} %

\title{\MyTitle}

\author{\MyAuthora{}} %
\email{scott.genin@otilumionics.com} %
\affiliation{OTI Lumionics Inc.,
  100 College St. \#351, Toronto, Ontario\, M5G~1L5, Canada} %

\author{\MyAuthorb{}} %
\affiliation{OTI Lumionics Inc., 100 College St. \#351, Toronto,
  Ontario\, M5G~1L5, Canada} %

\author{\MyAuthorc{}} %
\affiliation{Department of Chemistry, The University of British
  Columbia, 2036 Main Mall, Vancouver, British Columbia\, V6T~1Z1,
  Canada} %

\author{\MyAuthord{}} %
\affiliation{OTI Lumionics Inc., 100 College St. \#351, Toronto,
  Ontario\, M5G~1L5, Canada} %

\author{\MyAuthore{}} %
\affiliation{OTI Lumionics Inc., 100 College St. \#351, Toronto,
  Ontario\, M5G~1L5, Canada}

\author{\MyAuthorf{}} %
\email{zhudson@chem.ubc.ca} %
\affiliation{Department of Chemistry, The University of British
  Columbia, 2036 Main Mall, Vancouver, British Columbia\, V6T~1Z1,
  Canada} %

\begin{document}


\begin{abstract}
  Quantum chemistry simulations that accurately predict the properties
  of materials are among the most highly anticipated applications of
  quantum computing. It is widely believed that simulations running on
  quantum computers will allow for higher accuracy, but there has not
  yet been a convincing demonstration that quantum methods are
  competitive with existing classical methods at scale. Here we apply
  the \gls{iQCC} method on classical hardware to the calculation of
  the T$_1$ → S$_0$ transition energies in nine phosphorescent iridium
  complexes, to determine if quantum simulations have any advantage
  over traditional computing methods. Phosphorescent iridium complexes
  are integral to the widespread commercialization of \gls{OLED}
  technology, yet accurate computational prediction of their emission
  energies remains a challenge. Our simulations would require a
  gate-based quantum computer with a minimum of 72 fully-connected and
  error-corrected logical qubits. Since such devices do not yet exist,
  we demonstrate the \gls{iQCC} quantum method using a special purpose
  quantum simulator on classical hardware. The results are compared to
  a selection of common \gls{DFT} functionals (B3LYP, CAM-B3LYP,
  LC-wHPBE), \textit{ab initio} methods (HF and MP2), and experimental
  data. The \gls{iQCC} quantum method is found to match the accuracy
  of the fine-tuned \gls{DFT} functionals, has a better Pearson
  correlation coefficient, and still has considerable potential for
  systematic improvement. Based on these results, we anticipate that
  the \gls{iQCC} quantum method will have the required accuracy to
  design organometallic complexes when deployed on emerging quantum
  hardware.
\end{abstract}

\glsresetall

\maketitle

\section{Introduction}
\label{sec:introduction}

Computational chemistry methods such as quantum chemistry simulations
are becoming increasingly important tools for chemical, material, and
molecular design. Whether simulating molecular docking in
pharmaceuticals or the optical gaps of semiconductors, simulations
underpin several multi-trillion-dollar chemical and material
industries~\cite{Marzari:2021/natmat/736, Jorgensen:2004/sci/1813,
  Liu:2017/pnas/6990, Ackland:2017/sci/1254}. Results from simulations
are typically either used to screen potential candidates for desired
properties, or used to train predictive machine learning
models~\cite{Nandy:2018/iecr/13973,
  Joung:2021/jacs_au/427,Moosavi:2020/jacs/20273,
  Fanourgakis:2020/jacs/3814}. In either case, robust and accurate
quantum chemistry methods are needed to reliably predict
structure-property relationships in order to progress from use in
rough screening towards precision \textit{in silico} design. However,
the exact solution to the electronic Schrödinger equation, known as
the \gls{FCI}, scales exponentially with system size and is thus
intractable for all but the smallest molecules. To resolve this issue,
many approximate methods have been developed that are either
\textit{ab initio} or semi-emperical. \Gls{DFT} is one of the most
used types of quantum chemistry simulation, but often requires
fine-tuning of the functionals to achieve a high level of accuracy.
Furthermore, \gls{DFT} often performs poorly for more complex systems
with higher degrees of electron correlation, particularly when trying
to simulate excited or transition states~\cite{Becke:2014/jcp/18A301}.
New quantum chemistry methods capable of accurately simulating
electron correlation energies without first tuning parameters for a
specific problem are thus highly desirable and are an active area of
research.

Quantum computers, which operate using quantum states of matter, have
recently attracted major research interest as potential new tools for
high-accuracy quantum chemistry
simulations~\cite{Feynman:1982/ijtp/467, Lloyd:1996/sci/1073,
  Abrams:1997/prl/2586, Ladd:2010/nature/45}. The largest quantum
chemistry simulation to date on a quantum computer was limited to a
Hartree--Fock simulation of \ce{H_{12}} and diazene using 12 logical
qubits~\cite{GoogleAI:2020/sci/1084}, problems that are otherwise
straightforward to simulate on a classical computer. It is not yet
possible to perform a quantum chemistry simulation that is difficult
(or impossible) to solve on a classical computer, using current
\gls{NISQ} computers. It has been postulated that a quantum computer
with 50--100 logical qubits (or $\sim 10^2$--$10^4$ physical qubits
depending on the error correction
scheme~\cite{Fowler:2012/pra/032324}) would provide quantum advantage
over a classical computer for quantum chemistry
problems~\cite{Arute:2019/nat/505, Elfving:2020/arxiv/2009.12472}. At
this scale, a quantum computer could potentially enable accurate
quantum chemistry simulations of small molecules that are large enough
to be commercially relevant in a variety of industries, and that are
otherwise known to be difficult to simulate with traditional methods,
such as organometallic complexes.

The electronic structure of late transition metal complexes are known
to be difficult to accurately simulate due to the large number of
orbitals, the challenge in modeling electron correlation, and the
relativistic effects of the metal centre~\cite{Nandy:2021/cr/9927}.
Nonetheless, these complexes are important to multiple industries, for
example as catalysts, semiconductor precursors, pharmaceuticals and as
phosphorescent emitters for organic \glspl{OLED}. Cyclometalated
iridium (III) complexes are common emitters used commercially in
\gls{OLED} displays and lighting, due to their high quantum yield and
short radiative triplet lifetimes~\cite{Lamansky:2001/jacs/4304,
  Sajoto:2009/jacs/9813, Lai:2018/jacs/10198, You:2009/dtran/1267}.
The phosphorescent emission colors of these complexes are determined
by the triplet-to-singlet (T$_1$ → S$_0$) transition, making this an
ideal problem for testing new quantum chemistry simulations, since
these transitions i) can easily be measured experimentally with high
reproducibility; ii) require the simulation of both ground and excited
states, and iii) have been extensively studied using
\gls{DFT}~\cite{Hohenberg:1964/pr/b864, Kohn:1965/pr/a1133} and
\gls{TDDFT}~\cite{Runge:1984/prl/997, Bowler:2010/jp_cond/074207}.
These transitions are known to be difficult problems to simulate, as
\gls{DFT} methods commonly have an error of
$\sim \SI{0.3}{\electronvolt}$~\cite{Horbatenko:2019/jpca/7991}, which
is sufficient as a rough screening tool, but not for precision
\textit{in silico} design. While \gls{DFT} functionals can be
fine-tuned for a particular system to improve accuracy, choosing an
optimal functional is a daunting task due to the number of functionals
available.

Unfortunately, at the time of writing, a quantum computer with enough
qubits and sufficient error rate suitable for simulating the T$_1$ →
S$_0$ transition in phosphorescent emitters does not yet exist.
Nonetheless, to test if quantum chemistry simulations on a quantum
computer could provide any advantage for this class of problems, we
performed simulations using the \gls{iQCC} quantum method on a
classical computer~\cite{Ryabinkin:2018/jctc/6317,
  Ryabinkin:2020/jctc/1055, Ryabinkin:2021/qst/024012}. The \gls{iQCC}
method formulates the electronic structure problem natively in qubit
space in a highly efficient manner and was originally designed to
overcome many of the hardware limitations of today’s \gls{NISQ}
quantum computers. As a result of \gls{iQCC}’s highly efficient
formulation, problems that would otherwise require too many quantum
gate operations to practically solve on a classical computer can still
be solved using modest classical hardware. In effect, our \gls{iQCC}
quantum-on-classical-hardware simulations are emulating the
capabilities of a future quantum computer using classical hardware
available today. This approach has the advantage of having full qubit
connectivity and no error, while also facilitating large-scale
calculations that cannot be run on currently available quantum
computers. Using the \gls{iQCC} quantum simulator, we evaluate whether
a quantum computer could accurately simulate the T$_1$ → S$_0$
transition in phosphorescent emitters and whether this approach has
any advantages to existing classical methods such as \gls{DFT}.

\section{Results and discussion}
\label{sec:results-discussion-1}

\begin{figure}
  \centering \includegraphics[width=0.5\textwidth]{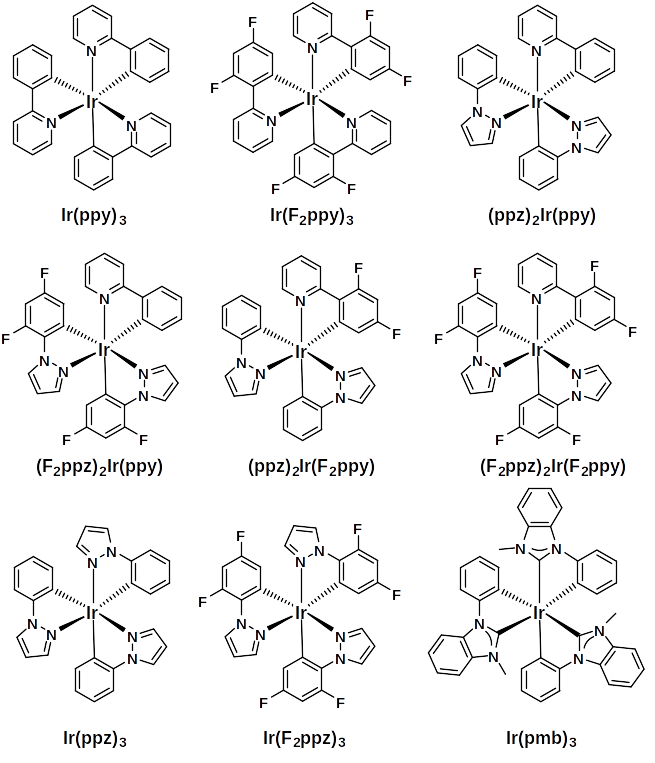}
  \caption{The nine cyclometalated \ce{Ir (III)} complexes simulated
    here.}
  \label{fig:1}
\end{figure}

\begin{figure}
  \centering \includegraphics[width=0.5\textwidth]{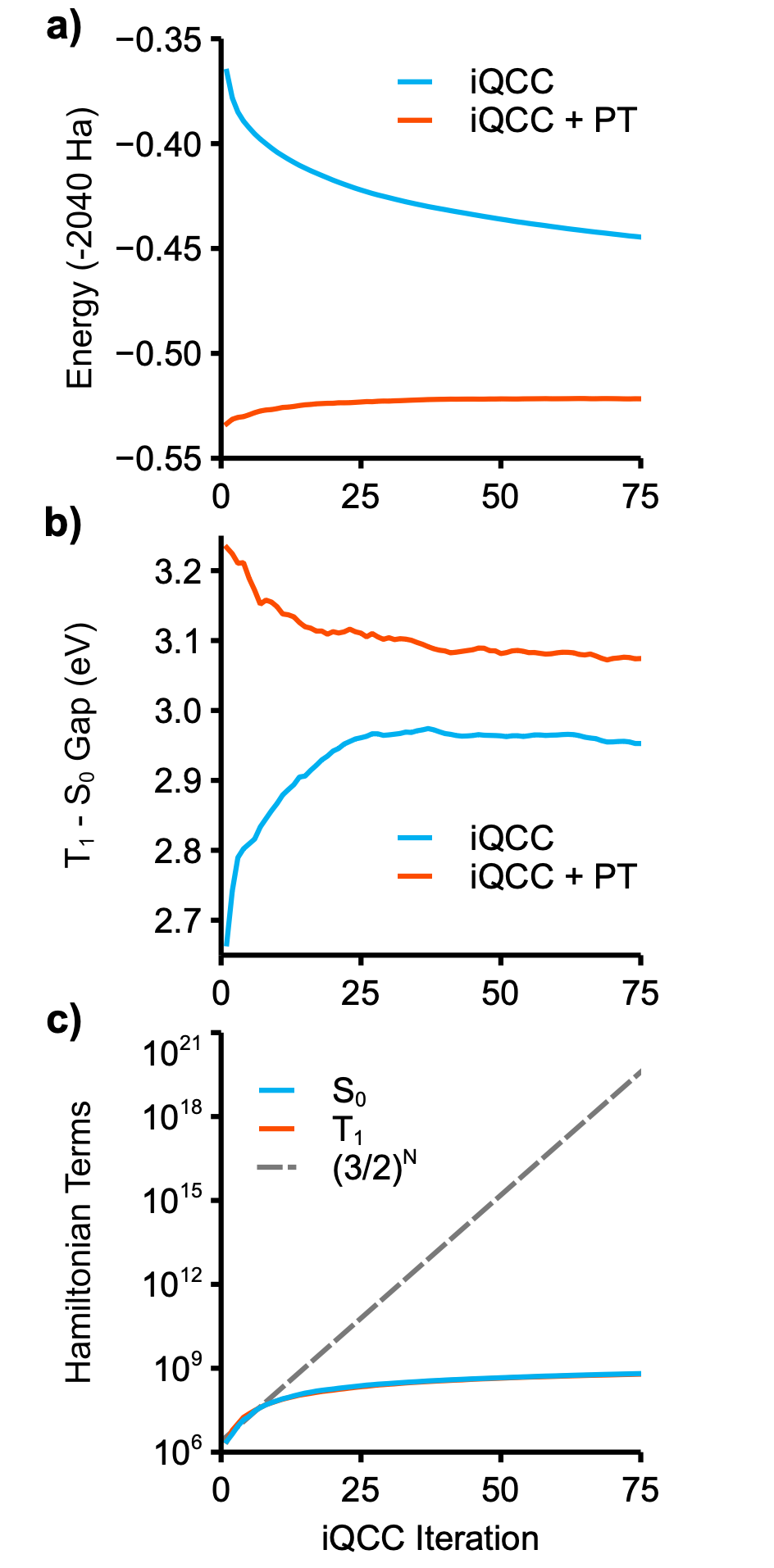}
  \caption{Representative convergence curves for \ce{Ir(pmb)_3} with
    \gls{iQCC} and \gls{iQCC} + PT \si{\hartree} for the S$_0$ state
    (scale minimum is \SI{-2040.55}{\hartree}), b) T$_1$-S$_0$ gap in
    \si{\electronvolt} plotted per \gls{iQCC} iteration for both
    \gls{iQCC} and \gls{iQCC} + PT, and c) Hamiltonian term size with
    respect to \gls{iQCC} iteration for both S$_0$ and T$_1$ for
    \ce{Ir(pmb)_3}, including a line to show the theoretical
    Hamiltonian term growth.}
  \label{fig:2}
\end{figure}

The geometry of nine cyclometalated \ce{Ir (III)} complexes, see
Fig.~\ref{fig:1}
were first optimized at T$_1$ state using the range-separated
\gls{DFT} functional, CAM-B3LYP~\cite{Yanai:2004/cpl/51} with the
LANL2def-DZV basis set~\cite{Hay:1985/jcp/299, Roy:2008/jctc/1029}
with unrestricted Hartree--Fock orbitals. Singlet and triplet states
were subsequently simulated using several \textit{ab initio},
\gls{DFT}, and \gls{TDDFT} simulations. The \textit{ab initio}
low-scaling Hartree--Fock and \gls{MP2} methods were used, as well as
well-known \gls{DFT} functionals (B3LYP)~\cite{Becke:1993/jcp/5648}
and range-separated \gls{DFT} functionals (CAM-B3LYP and
LC-wHPBE)~\cite{Vydrov:2006/jcp/234109} with both LANL2def-dzv and
SBKJC-ECP 6-31G* basis sets~\cite{Stevens:1992/cjc/612}. \Gls{TDDFT}
simulations of the triplet states with the same functionals were run
with and without the Tamm-Dancoff
approximation~\cite{Peach:2011/jctc/3578}. The optimized triplet state
geometries were used to carry out spin-restricted open-shell
Hartree--Fock \gls{MO} calculations to generate the values of one- and
two- electron integrals using a modified version of
GAMESS.~\cite{Gamessus, Gamessus-2} 18 occupied and 18 unoccupied
Hartree--Fock \gls{MO}s below and above the Fermi level constitute a
\gls{CAS} containing 36 active electrons over 36 active spin orbitals.
The one- and two-electron integrals from the CAS(36,36) were used to
construct the qubit Hamiltonian for \gls{iQCC} using the Jordan-Wigner
transformation~\cite{Nielsen:2005/scholar_text}. The \gls{iQCC}
quantum simulator was used to solve for the converged T$_1$ and S$_0$
energies applying corrections from perturbation theory
(PT)~\cite{Ryabinkin:2018/jctc/6317, Ryabinkin:2020/jctc/1055,
  Ryabinkin:2021/qst/024012} (Fig.~\ref{fig:2}). Here, the PT
correction refers specifically to Epstein-Nesbet perturbation theory,
which applies a perturbative correction to the \gls{iQCC}
energy~\cite{Ryabinkin:2021/qst/024012}. These results are equivalent
to those that would be produced using a quantum computer with 72
\textit{logical} qubits, and which is within the range of problem size
where a quantum advantage has been
postulated~\cite{Yuan:2020/science/1054}.
 
The \gls{iQCC} method has been improved upon since its disclosure in
earlier reports, as described in the supplementary information. The
\gls{iQCC} method was originally developed to enable flexibility in
quantum circuit depth by allowing an Ansatz, or quantum instruction
set, to be partitioned and optimized
independently~\cite{Ryabinkin:2020/jctc/1055}. The Ansatz is then
incorporated or ``dressed'' into the Hamiltonian for the next step (SI
Eq.~16). Since the Ansatz does not fully commute with the Hamiltonian,
the incorporation results in additional terms being added into the
Hamiltonian, with exponential growth proportional to $(3/2)^N$, but in
every case the growth plateaued, enabling efficient scaling of the
\gls{iQCC} quantum simulator on classical hardware for these problems
(Fig.~\ref{fig:2}c).
  
\begin{figure}
  \centering \includegraphics[width=1.0\textwidth]{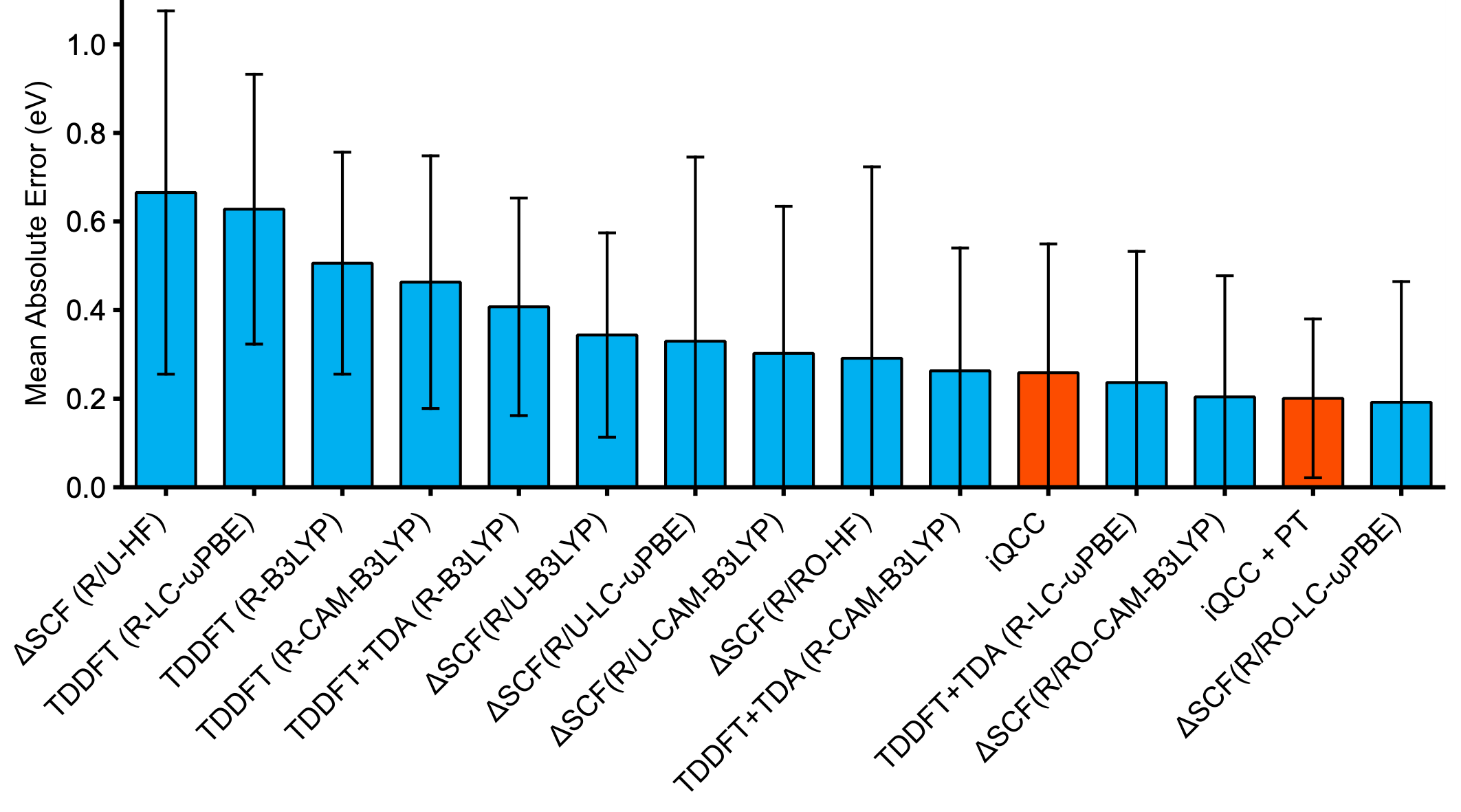}
  \caption{Mean absolute deviations from experimental data for the
    different electronic structure methods. Error bars indicate the
    standard deviation. Experimental T$_1$-S$_0$ energies were
    obtained at \SI{77}{\kelvin} in dilute frozen \ce{2-MeTHF}
    solution~\cite{Sajoto:2009/jacs/9813}.}
  \label{fig:3}
\end{figure}

The mean accuracy of the \gls{iQCC}+PT simulations for the set of Ir
complexes was superior to nearly all of the studied \gls{DFT} and
\textit{ab initio} methods for simulating the T$_1$ → S$_0$ transition
for this set of complexes. Only the fine-tuned LC-wHPBE functionals
produced energies with lower mean absolute error, although with higher
standard deviation (Fig.~\ref{fig:3}). All methods tested showed
variations from the experimental value, attributable to the
approximations and/or intrinsic errors inherent to each method. The
absence of solvation effects and the use of double-zeta basis sets are
likely the two largest sources of error across all methods. The
current limitations of the \gls{iQCC} and \gls{iQCC}+PT results can be
attributed to the limited \gls{CAS} size used, which was selected for
practical consideration. At this size, the total number of all
possible electronic configurations (of any spin) is
${36 \choose 18}^2 \approx 10^{20}$. Apparently, a problem too large
for using the \gls{FCI} method on a classical computer. The CAS
limitation results in the treatment of electrons and orbitals outside
of the CAS only at the mean-field level, so there is an implicit
assumption the electron correlation outside of the CAS is not
significant.

The best-performing \gls{DFT} functionals for this set were the
long-range corrected functionals LC-wHPBE and CAM-B3LYP when using
restricted open shell orbitals, however when using unrestricted
orbitals, the B3LYP, CAM-B3LYP and LC-wHPBE had very similar average
values. The common \gls{DFT} functional PBE was also tested using
restricted open shell orbitals and showed a much larger absolute
deviation (1.437 eV compared to $\sim \SI{0.2}{\electronvolt}$ for
CAM-B3LYP and LC-wHPBE) and is thus not shown below. There was also a
large range in error for the \textit{ab initio} methods; while
\gls{MP2} is often considered to be more accurate than HF, this was
not the case here (SI table). The PT correction introduced a blue
shift to the T$_1$ → S$_0$ transition on average of
\SI{0.3}{\electronvolt} eV with a range of 0.04-0.72 eV, resulting in
7 of the 9 Ir (III) complexes examined showing a blue shift in
phosphorescence emission relative to experiment. This outcome was
expected based on the absence of solvent and thermal effects in the
simulations, which would be expected to red-shift the experimental
values.

To be useful as a precision \textit{in silico} design tool, it is
important that a method can correctly predict structure-property
relationships, or in other words the correlation between changes in
chemical structure and changes in electronic structure. It was found
that the \gls{iQCC} and \gls{iQCC}+PT results have higher Pearson’s
correlation coefficients than even those using the LC-wHPBE
functional, and that the \gls{iQCC}+PT is significant at the 95\%
confidence interval (Table~\ref{tab:1}). This implies that
\gls{iQCC}+PT is more accurate at predicting the structure-property
relationship and could potentially be a more useful tool for chemical
design compared to the best-performing \gls{DFT} methods examined.

\begin{table}
  \centering
  \caption{Mean average deviation (\si{\electronvolt}), standard
    deviation (\si{\electronvolt}), Pearson’s correlation coefficients
    (higher is better) and P-values (closer to 0 is better) for the
    best performing methods.}
  \begin{minipage}{1.0\linewidth}
    \sisetup{ %
      table-format=4.4}
    \begin{tabularx}{1.0\linewidth}{lSSSS}
      \toprule \\
      Method & \multicolumn{1}{p{2,7cm}}{Mean average deviation, \si{\electronvolt}} & \multicolumn{1}{p{2.4cm}}{Standard \mbox{deviation}, \si{\electronvolt}} &\multicolumn{1}{l}{Pearson} & \multicolumn{1}{c}{P-value} \\
      \midrule
      ROHF/RHF &                  0.291 & 0.432 & -0.413 & 0.269 \\
      ROMP2/RMP2\footnotemark[1]& 0.246 & 0.390 &  0.438 & 0.276 \\
      RO/R LC-wHPBE &             0.192 & 0.272 &  0.328 & 0.388 \\
      TDDFT+TDA, LC-wHPBE &       0.236 & 0.296 &  0.255 & 0.507 \\
      iQCC &                      0.258 & 0.291 &  0.522 & 0.149 \\
      iQCC + PT &                 0.201 & 0.179 &  0.762 & 0.018 \\
      \bottomrule
    \end{tabularx}
  \end{minipage}

  \raggedright %
  \footnotetext{1}{\footnotesize $^1$ An outlier data point was
    removed for R/RO-MP2 for \ce{Ir(ppy)_3}, gap =
    \SI{8.9}{\electronvolt}}
  \label{tab:1}
\end{table}

In summary, nine cyclometalated Ir (III) complexes were calculated
using the \gls{iQCC} method using a special purpose quantum simulator
that emulates a quantum computer with 72 logical qubits. The results
are found to be competitive with, or superior to, many leading
\gls{DFT} and \textit{ab initio} methods, when compared to
experimental data. Specifically, the \gls{iQCC} method combined with
PT corrections is found to match the accuracy of fine-tuned \gls{DFT}
functionals for simulating the T$_1$ → S$_0$ transition, but with more
accurate prediction of the structure-property relationship. This
implies that the \gls{iQCC} quantum method has the potential to be
robust and reliable enough to simulate a set of materials without
parameter tuning, which will be beneficial to the design of \textit{de
  novo} materials when experimental data is not widely available.
These results show that a quantum computer with fully connected and
error-corrected qubits will be capable of accurately simulating
chemical properties, and provides an optimistic view on the future of
quantum methods for chemical design. Finally, the \gls{iQCC}
quantum-on-classical-hardware simulations demonstrated here are not
limited to just simulation of the T$_1$ → S$_0$ transition in
phosphorescent emitters and has already been successfully used to
simulate other chemical systems and properties, the results of which
will be published elsewhere.

\bibliography{matsim}

\end{document}


\beginsupplement

\section{The \gls{iQCC} method}
\label{sec:overv-iqcc-algor}

\subsection{Second-quantized electronic Hamiltonian}
\label{sec:second-quant-electr}

The \gls{QCC} method starts from a second-quantized electronic
Hamiltonian of a molecule~\cite{Helgaker:2000, Abrams:1997/prl/2586,
  AspuruGuzik:2005/sci/1704, Peruzzo:2014/ncomm/4213}:
\begin{equation}
  \label{eq:qe_ham}
  \hat H_e = \sum_{ij}^{N_\text{so}} h_{ij} {\hat a}^\dagger_i {\hat
    a}_j + \frac{1}{2}\sum_{ijkl}^{N_\text{so}} 
  g_{ijkl} {\hat a}^\dagger_i {\hat a}^\dagger_j {\hat a}_l {\hat a}_k,
\end{equation}
where ${\hat a_i}^\dagger$ and ${\hat a_i}$ are fermion creation and
annihilation operators, and
\begin{eqnarray}
  \label{eq:1eints}
  h_{ij} = & \int \psi_i^{*}(\mathbf{x})\left(-\frac{1}{2}\nabla^2 - \sum_I
             \frac{Z_I}{|r-R_I|}\right) \psi_j(\mathbf{x}) d\mathbf{x}, \\
  \label{eq:2eints}
  h_{ijkl} = & \int\int \dfrac{\psi_i^{*}(\mathbf{x}_1)\psi_k^{*}(\mathbf{x}_2)\psi_j(\mathbf{x}_1)\psi_l(\mathbf{x}_2)}{r_{12}} \mathrm{d}\mathbf{x}_1 \mathrm{d}\mathbf{x}_2
\end{eqnarray}
are values of one- and two-electron integrals written in a
spin-orbital basis of $\psi_j(\mathbf{x})$;
$\mathbf{x} = (\mathbf{r}, \sigma)$. They were computed using the
locally modified \textsc{gamess-us}~\cite{gamessus, gamessus-2}
[ver.~30 SEP 2019 (R2)] program.

The electronic Hamiltonian~\eqref{eq:qe_ham} is converted to a qubit
form by the \gls{JW}
transformation~\cite{Jordan:1928/zphys/631,AspuruGuzik:2005/sci/1704}
to obtain
\begin{equation}
  \label{eq:qubit_H}
  \hat H = \sum_{k=1}^{M} C_k \hat P_k,
\end{equation}
where $C_k$ are coefficients inferred from the molecular integrals,
and $\hat P_k$ are Pauli strings (``words''), the products of Pauli
elementary operators $\sigma_j = x_j$, $y_j$, or $z_j$ acting on an
individual $j$-th qubit,
\begin{equation}
  \label{eq:P_k-def}
  P_k = \prod_{j \ge 0,\ j \in j(k)} \sigma_j.
\end{equation}
As long as the molecular Hamiltonian is a \emph{real} operator, which
is always the case in the absence of a magnetic field, every $P_k$
contains the \textit{even} number of $y_j$ factors.

The pairwise grouping of spin-orbitals was employed: the first orbital
with $\alpha$ (up) spin is mapped to the first qubit, followed by the
first orbital with $\beta$ spin, which is mapped to the second, etc.
The advantage of using \gls{JW} transformation is one-to-one
correspondence between population of fermion spin-orbitals and states
of qubits. We introduce a qubit reference vector
\begin{align}
  \label{eq:qubit_ref}
  \ket{0} & = \prod_{k=1}^{n_e} \ket{\downarrow}_k \times \prod_{k=1}
            ^{N-n_e} \ket{\uparrow}_k,
\end{align}
where $n_e$ is the number of electrons (equals to the number of
\emph{occupied} spin-orbitals), and $N$ is the total number of qubits.
Hence, $(N-n_e)$ qubits correspond to \emph{unoccupied} (virtual)
orbitals. Together, employed occupied and virtual spin-orbitals
constitute the \gls{CAS} for the problem. If $N$ is twice the size of
the atomic basis, one deals with the \gls{FCI} problem, otherwise it
is a \gls{CASSCF}-like problem with the active space which is
conveniently abbreviated as CAS($n_e$, $N$). It must be noted that the
\gls{iQCC} method with \gls{CAS} other than the all electron/full
basis (the \gls{FCI} problem) is \emph{not} equivalent to either the
conventional \gls{CASSCF} (which performs the full orbital
optimization) or the \gls{CASCI} (which does not touch orbitals at
all) problem. \gls{iQCC} does not distinguish between orbital and
CI-coefficient optimizations within the active space, but does not
change orbitals outside \gls{CAS}.

The state given by Eq.~\eqref{eq:qubit_ref} represents a simple qubit
product state---a special case of previously used qubit coherent
states~\cite{Ryabinkin:2018/jctc/6317}. The new feature of the current
\gls{iQCC} implementation is that this state is fixed throughout
\gls{iQCC} iterations, see Sec.~\ref{sec:vqe-style-energy} for
details.

\subsection{Normal-ordered qubit Hamiltonians}
\label{sec:normal-ordered-qubit}

The qubit Hamiltonian, Eq.~\eqref{eq:qubit_H}, is a simple linear
combination of Pauli words. Although this form is general, it
does not offer any advantage for evaluation of the quantum-mechanical
expectation values. It can be regrouped, however, as
\begin{equation}
  \label{eq:Ising_decomp}
  \hat H = {\hat I}_0(\mathbf{z}) + \sum_{k >0} {\hat I_k}(\mathbf{z}) X_k,
\end{equation}
which we call the ``Ising decomposition'', and which is convenient to
set up the \gls{iQCC} method. ${\hat I_k}(\mathbf{z})$ for
$k = 0, 1, \dots$ are qubit Hamiltonians which are the sum of Pauli
words containing only $\hat z$ Pauli elementary operators
(``generalized Ising Hamiltonians''). All $X_k$ are the Pauli
$X$-strings. Namely,
\begin{equation}
  \label{eq:X-strings}
  X_k = \prod_{j \ge 0,\ j \in j(k)} \hat x_j.
\end{equation}
To evaluate the mean values with operators in their Ising
decomposition, we need a qubit reference vector [see
Eq.~\eqref{eq:qubit_ref}], and action rules for elementary Pauli
operators $\hat z_i$:
\begin{align}
  \label{eq:z_action}
  \hat z_i \ket{\downarrow}_j & = (1 - 2\delta_{ij})
                                \ket{\downarrow}_j \\
  \hat z_i \ket{\uparrow}_j   & = \ket{\uparrow}_j
\end{align}
Note that any $X$-string operators have always \emph{zero} matrix
elements, $\braket{0|X_k|0} = 0$ for any $\ket{0}$, so that they are
purely off-diagonal operators. The following identity holds (taking
$\hat H$ as an example):
\begin{equation}
  \label{eq:H_00}
  \braket{0|\hat H|0} = \braket{0|{\hat I}_0|0},  
\end{equation}
where ${\hat I}_0$ is the corresponding operator from the Ising
decomposition~\eqref{eq:Ising_decomp}.

\subsection{The \acrfull{QCC} Ansatz}
\label{sec:acrfullqcc-ansatz}

In the current study we use a generic \gls{QCC} form:
\begin{equation}
  \label{eq:U_QCC}
  \hat U(\mathbf{t}) = \hat U(t_1, \dots, t_L) =
  \prod_{j=1}^L\exp\left(-\I t_j \hat T_j/2\right),
\end{equation}
where $t_j$ are amplitudes to be optimized, $\hat T_j$ are generators,
and $L \le 16$ is the number of them. The upper limit is dictated by
the current implementation (see Sec.~\ref{sec:vqe-style-energy}) and
the hardware resources available; in general, computation costs
increase \emph{exponentially} with $L$.

\subsection{\gls{QCC} operator ranking}
\label{sec:glsqcc-oper-rank}

$\hat T_k$, the generators of the \gls{QCC} form, can be derived from
$X_k$ operators appearing in the Ising
decomposition~\eqref{eq:Ising_decomp}\cite{Ryabinkin:2021/qst/024012}.
In particular, to convert a given $X_k$ into $T_k$ that is
characterized by a non-zero value of the energy gradient
\begin{equation}
  \label{eq:energy_grad}
  \frac{\mathrm{d}E[T_k]}{\mathrm{d}t}\Bigg|_{t_k=0} =
  \frac{1}{2\I}\Braket{0|[\hat H, \hat T_k]|0} =
  \frac{1}{2}\Im{\Braket{0|[\hat H, \hat T_k]|0}}, 
\end{equation}
one has to substitute the \emph{odd} number of elementary $\hat x_j$
operators in $\hat X_k$ for their $\hat y$ counterparts. Only such
generators may have non-zero gradients and hence, change the total
energy as they are purely \textit{imaginary} operators. We always
replace a single $\hat x_j$ with the smallest possible $j$ and call
that choice ``canonical''.

The energy expectation value at the $i$-th \gls{iQCC} iteration,
\begin{equation}
  \label{eq:QCC_energy_expr}
  E^{(i)}(t_1, \dots, t_L) = \Braket{0|\left(U^{(i)}\right)^\dagger(t_1,
    \dots, t_L) \hat H^{(i-1)} U^{(i)}(t_1, \dots, t_L)|0},
\end{equation}
depends on the ordering of generators in Eq.~\eqref{eq:U_QCC}. We sort
all possible generators [whose number is equal to the number of terms
in the sum in the right-hand side of Eq.~\eqref{eq:Ising_decomp}]
according to their ``importance measure''. It can be an absolute
gradient value, Eq.~\eqref{eq:energy_grad}, the choice that were used
in Ref.~\citenum{Ryabinkin:2020/jctc/1055}, or an absolute value of
the optimal amplitude~\cite{Ryabinkin:2021/qst/024012} -- the current
choice. For the sake of completeness, we derive the formula for the
latter below. Consider the \gls{QCC} form with a single generator
$\hat T$:
\begin{eqnarray}
  \label{eq:dEdT}
  E[\hat T](t)  & = & \Braket{0|\exp(\I t\hat T/2) \hat H \exp(-\I t\hat T/2)|0} \nonumber \\
           & = &  \braket{0|\hat H|0} + \frac{\sin(t)}{2\I}
                 \Braket{0|[\hat H, \hat T]|0} +
                 \frac{1 - \cos(t)}{2} \Braket{0|(\hat T \hat H \hat T
                 - \hat H)|0}  
\end{eqnarray}
Identifying~\cite{Ryabinkin:2021/qst/024012}
\begin{eqnarray}
  \label{eq:omegas}
  \omega & = & \frac{1}{2}\left|\braket{0|[\hat H, \hat T]|0}\right| =
               \left|\braket{0|\hat H \hat T|0}\right| \\
  \label{eq:denom}
  D & = & \Braket{0| \hat T \hat H \hat T - \hat H | 0}
\end{eqnarray}
we can write:
$E(t) = E_0 + \omega \sin\left(t\right) + D\left(1 - \cos t
\right)/2$, which gives the energy lowering of
$\Delta E = D/2 - \sqrt{\left(D/2\right)^2 + \omega^2} \approx
-\frac{\omega^2}{D} < 0$ if $D > 0$, when
\begin{equation}
  \label{eq:EN1_amplitude_formula}
  t = \arcsin\frac{2\omega}{\sqrt{D^2 + 4\omega^2}}.
\end{equation}
If $\hat T \equiv \hat T_k$ for some generator (\emph{e.g.} canonical)
derived from the Ising decomposition as described above, then the
corresponding $|t_k|$ is the sought-for importance measure.
Additionally,
\begin{equation}
  \label{eq:omega_k}
  \omega_k = \left|\braket{0|\hat I_k|0}\right|,
\end{equation}
where $\hat I_k$ is taken from the Ising decomposition of the current
Hamiltonian, Eq.~\eqref{eq:Ising_decomp}.

When $T_k$ are sorted in descending order (highest come first)
according to the chosen importance measure, top $L$ of them are
selected for the \gls{QCC} Ansatz, Eq.~\eqref{eq:U_QCC}, for the next
iteration. The remaining generators are used to set up the
perturbative correction, see Ref.~\citenum{Ryabinkin:2021/qst/024012}
for details.

\subsection{\Gls{VQE}-style energy optimization}
\label{sec:vqe-style-energy}

Optimization of the \gls{QCC} energy
expression~\eqref{eq:QCC_energy_expr} is a non-linear optimization
problem. We solved it on a classical computer with the aid of
\textsc{NL}opt optimization library~\cite{NLOpt}. Specifically, the
low-storage BFGS algorithm\footnote{Implemented and posted online as a
  LGPL Fortran library by Ladislav Luksan~\cite{LD_BFGS}. The L-BFGS
  method itself is originated from Refs.~\citenum{Nocedal:1980/mp/773}
  and~\citenum{Liu:1989/mp/503}.} was employed. The non-linear energy
functional~\eqref{eq:QCC_energy_expr} was implemented as a part of our
quantum simulator; it is capable of evaluating energy values and
corresponding gradients\footnote{On real quantum hardware gradients
  can be efficiently evaluated using the parameter shift
  approach~\cite{Schuld:2019/pra/032331}}.

Eq.~\eqref{eq:EN1_amplitude_formula} was used to provide the initial
guess to amplitudes with $\omega_k$ evaluated (along with correct
signs) by Eq.~\eqref{eq:omega_k}. With such a choice of the initial
amplitudes, the number of the objective function evaluations during
optimization was less than 15. No additional constraints were applied,
but the triplet energy functional was constructed by adding a penalty
term
\begin{equation}
  \label{eq:triplet_penalty}
  \mu \hat W = \mu \left(\hat S^2 - s(s+1)\hat S_z\right),
\end{equation}
where $\hat S^2$ and $\hat S_z$ are the total spin squared operator
and its $z$ projection, respectively, $s = 1$ is a spin quantum number
for a triplet state, and $\mu = 0.25$ is a penalty parameter, to the
Hamiltonian,
\begin{equation}
  \label{eq:penalized_ham}
  \hat H \to \hat H + \mu \hat W.
\end{equation}
We also added the penalty term with the same $\mu$ and $s=0$ for the
singlet state to obtain more correct gap values as the energy minima
for penalized operators are close, but not identical to those for an
un-penalized problem~\cite{Nocedal:2006}.

Contrary to the previous studies, (\emph{e.g.}
Ref.~\citenum{Ryabinkin:2020/jctc/1055}), single-qubit (mean-field)
optimizations were not performed, and the qubit reference state
$\ket{0}$ was fixed during \gls{iQCC} iterations. The rationale was
the following: the \gls{QCC} form is flexible enough to include
generators of a single-qubit transformations, which are simply
$\hat T_k = \hat y_k$. Whenever such generators are necessary (which
is determined by their large values of the important measure), they
will be included into the \gls{QCC} Ansatz automatically.

\subsection{Hamiltonian dressing}
\label{sec:hamiltonian-dressing}

Once the optimal values of amplitudes $\mathbf{t}_\text{opt}$ are
found, the \gls{iQCC} program performs \emph{dressing} of the current
Hamiltonian. Below we show an elementary step of this transformation,
which corresponds to $L=1$ in Eq.~\eqref{eq:U_QCC}. For $L > 1$ these
steps are performed recursively. Dressing is a unitary transformation
of the current Hamiltonian using the \emph{optimal numerical} values
of amplitude(s) at the current iteration:
\begin{align}
  \label{eq:dressing}
  \hat H^{(i+1)}  = \left(U^{(i)}\right)^\dagger(t_\text{opt}) \hat
  H^{(i)} U^{(i)}(t_\text{opt}) & = \hat H^{(i)} -
                                  \frac{\I}{2}\sin(t_\text{opt}) [\hat
                                  H^{(i)}, \hat T] + \frac{1 - \cos
                                  t_\text{opt}}{2}\left(\hat T \hat
                                  H^{(i)} \hat T - \hat
                                  H^{(i)}\right).
\end{align}
New terms in $\hat H^{(i+1)}$ come from the second, commutator term in
the right-hand side of Eq.~\eqref{eq:dressing}, because an operator
$(\hat T \hat H \hat T - \hat H)$ shares the same terms with $\hat H$
for \emph{any} Pauli word $\hat T$. ``On average'' an arbitrary
$\hat T$ commutes with a half of terms in $\hat H$, so after the
dressing the number of terms in $\hat H^{(i+1)}$ increases by a factor
of $3/2$~\cite{Ryabinkin:2020/jctc/1055}. In a more general case,
$L>1$, this translates into $(3/2)^L$, and for $K$ \gls{iQCC}
iterations into $(3/2)^{KL}$. Thus, \emph{formally}, the \gls{iQCC}
method is exponentially complex both in the size of the \gls{QCC}
Ansatz and the number of iterations. However, all new terms are
multiplied by some products of $\sin(t_\text{k,opt}) < 1$ for
$1 \le k \le L$. Additionally, for $L > 1$ $\sin(t_{k,\text{opt}})$
decrease with $L$ since $t_{k,\text{opt}}$ correlate with decreasing
values of the importance measure, see Sec.~\ref{sec:glsqcc-oper-rank}.
Also, \emph{maximal} $\sin(t_\text{opt})$ tends to decrease in the
course of \gls{iQCC} iterations. As a result, the dressed Hamiltonians
accumulate a lot of terms those numerical values are tiny, frequently
many orders of magnitudes smaller than the accuracy of the initial
Hamiltonian coefficients, which, in turn, is determined by the
accuracy of molecular integrals. Moreover, despite the abundance of
such terms their cumulative contribution to the ground-state energy is
negligibly small. This situation is somewhat known to the quantum
community; for example, Ref.~\citenum{Ivanic:2001/tca/339} coined the
term ``\gls{CI} deadwood''. Dropping such terms, however, not only
limits the growth (proliferation of terms) of dressed Hamiltonians,
but drastically changes the \emph{asymptotics} of such growth. As can
be seen from Fig.~2c of the main text, the number of numerically
\emph{significant} terms quickly deviates from the expected
exponential asymptotics. By fitting the Hamiltonian size as a function
of the iteration number, we found that at late iterations the actual
asymptotics is close to low-power polynomial, for example, $M^{1.608}$
for data on Fig~2c. This suggests that with our ``canonical'' choice
of generators, amplitudes of less than two generators are correlated,
and the second-order Epstein-Nesbet perturbation theory (see
Ref.~\citenum{Ryabinkin:2021/qst/024012}) may be very accurate.
Indeed, for our materials we observed that the most of the \gls{QCC}
top-ranked generators were originated from the initial (bare,
un-dressed) double fermionic excitations; rarely we saw single- and
higher-order composite excitations that results from dressing. This
implies, however, that the \gls{iQCC} method generates composite
(\textit{e.g.} high-order) excitations and combines them with the
original double fermionic excitations whenever necessary. In contrast
to other methods proposed in the literature, such as
ADAPT-VQE~\cite{Grimsley:2018/nc/3007} or a (generalized) \gls{UCC}
ones~\cite{Peruzzo:2014/ncomm/4213, Mcclean:2016/njp/023023,
  OMalley:2016/prx/031007, Romero:2018/qct/014008,
  Hempel:2018/prx/031022, Nam:2019/ArXiv/1902.10171,
  Lee:2019/jctc/311}, no preliminary selection of a ``generator pool''
or a generators' excitation rank is required to include energetically
important generators into Ans\"atze. The Hamiltonian itself naturally
drives the ground-state optimization, which allows to consider the
\gls{iQCC} method as a genuinely black-box electronic-structure
method.

It is also clear that for many molecular systems theoretical
(exponential) complexity of the iQCC procedure has little relevance to
the actual complexity, significantly widening the potential
applicability of the \gls{iQCC} method to realistic systems.

\section{Active orbital selection}
\label{sec:active-space-select-1}

For generating the triplet state Hamiltonians, we used the basis of
restricted open-shell Hartree--Fock \glspl{MO}. Namely, one- and
two-electron molecular integrals in Eq.~\eqref{eq:qe_ham} were
computed using the \gls{ROHF} orbitals. This choice was largely
unproblematic for all materials with an exception of \ce{Ir(ppz)_3}:
for this molecule we found an alternative set of \gls{ROHF} orbitals
with lower energy, namely, \SI{-1465.167381}{\hartree} \emph{vs.}
\SI{-1456.151315}{\hartree} found by \textsc{gamess} starting from the
``Huckel'' guess. New orbitals were generated by running the \gls{UHF}
calculations first; \gls{UHF} natural orbitals were prepared and used
subsequently as a guess for \gls{ROHF} calculations.

The space of active \glspl{MO} was chosen as follows: the lowest 18
occupied orbitals below the Fermi level and 18 virtual orbitals above
it were selected. In parallel \gls{UHF} calculations we found that 18
natural \gls{UHF} orbitals~\cite{Pulay:1988/jcp/4926} below the Fermi
level had occupation numbers within a range \numrange{1.998}{1.000},
while 18 above had occupation in a range \numrange{1.000}{0.0017}.
Hence, \gls{ROHF} orbitals that lie below HOMO-18 are likely to span
the same orbital space as their \gls{UHF} counterparts (because the
all have occupation numbers close to 2), and orbitals above constitute
the ``spin-polarized'' subsystem, which must be a part of the active
space~\cite{Abrams:2004/cpl/227}. An additional benefit of using
\gls{ROHF} orbitals is that, the initial \gls{QCC} energy matches the
\gls{ROHF} value if a qubit reference vector is chosen as in
Eq.~\eqref{eq:qubit_ref}.

Likewise, the choice of virtual orbitals for \gls{CAS} is rather
simple for the studied materials in the selected double-zeta basis set
(6-31G* basis set~\cite{Hehre:1972/jcp/2257} and its pseudopotential
partner for the \ce{Ir} center~\cite{Stevens:1992/cjc/612}). All
orbitals are ordered by their Hartree--Fock canonical energies,
$\epsilon_i < 0$ for the valence and $\epsilon_i > 0$ for the virtual
ones. In a double-zeta basis each occupied orbital is usually paired
with at least one virtual, so mechanical including lowest virtual
orbitals into the active space in general leads to a correct
description of correlation effects. If, for example, extended basis
sets with multiple \emph{diffuse} basis functions were used, there
would be numerous \emph{Rydberg} states, which are weakly bound (or
scattered) states with orbital energies close to 0 and having
virtually zero correlation contributions\footnote{Rydberg states can
  be largely avoided if additional (polarization) functions are
  systematically added by a procedure described in
  Ref.~\citenum{Rappoport:2010/jcp/134105}. It is demonstrated there
  that augmentation of a split-valence basis set with only one $d$
  function per atom leaves appreciable (23\%) basis set error.
  However, since we seek for the singlet-triplet energy
  \emph{difference}, error compensation seems to guarantees reasonably
  accurate differences in a chosen singly-augmented 6-31G* basis
  set.}. They would pollute the virtual space making the \gls{CAS}
selection difficult.

\section{Spin contamination of the UHF wave function}
\label{sec:spin-cont-uhf}

\begin{table}[t]
  \footnotesize %
  \centering
  \begin{tabularx}{0.73\linewidth}{@{}XS}
    \toprule
    Material & \multicolumn{1}{c}{Value} \\
    \midrule
    \ce{Ir(ppz)_2(F_2ppy)}      & 3.892 \\
    \ce{Ir(ppz)_3}              & 3.587 \\
    \ce{Ir(F_2ppz)_3}           & 2.412 \\
    \ce{Ir(F_2ppz)_2(F_2ppy)}   & 3.293 \\
    \ce{Ir(F_2ppz)_2(ppy)}      & 2.715 \\
    \ce{Ir(ppy)_3)}             & 4.879 \\
    \ce{Ir(ppz)_2(ppy)}         & 4.933 \\
    \ce{Ir(pmb)_3}              & 3.976 \\
    \ce{Ir(F_2ppy)_3}           & 4.411 \\
    \bottomrule
  \end{tabularx}
  \caption{$\Braket{\hat S^2}$ values for the \gls{UHF} determinant,
    an ideal value is $2$.}
  \label{tab:uhf-spin-contamination}
\end{table}
Notwithstanding what was said above about the electronic correlation
captured by the \gls{QCC} Ansatz, investigated molecules in their
triplet states \textit{cannot} be considered as weakly correlated.
Table~\ref{tab:uhf-spin-contamination} shows that the
single-determinant description of the triplet state is severely
flawed, showing strong spin contamination in the \gls{UHF} description
of those molecules.

\section{The Jahn-Teller effect in studied materials}
\label{sec:jahn-teller-effect}

All studied materials are susceptible to the
Jahn-Teller~\cite{Majumdar:2020/sc/7} effect in their triplet states.
Below we present a concise study of the Jahn-Teller effect in the
\ce{^3Ir(ppy)3} molecule~\cite{Fine:2012/mp/1849}. We have performed
\gls{CIS} calculations of triplet states in a geometrical
configuration of the \emph{singlet} state which has $C_{3}$ symmetry.
Three almost degenerate triplet states were found\footnote{$C_3$ group
  has no three-fold degenerate irreducible representations, hence, at
  least one degeneracy is either accidental or actually has a very
  small yet finite gap. This is not surprising since \emph{local}
  coordination of the \ce{Ir} atom is close to octahedral, and the
  $O_h$ group has there-fold degenerate irredicible representations.},
see Table~\ref{tab:Ir_ppy3_triplet_manifold}.

The Jahn-Teller effect leads to a geometrical distortion of the
\ce{^3Ir(ppy)_3} molecule and lifts the electronic degeneracy. As a
result, three energetically equivalent lower-symmetry minima emerge on
a largely flat, strongly anharmonic \gls{PES}. These minima are
connected by low barriers, so that the molecule is floppy and exhibit
large-amplitude motion even at low temperature. This also contributes
to the difficulties in geometry optimization as changing the level of
theory and various convergence thresholds lead to substantial changes
in the optimized structure. It appears that \emph{all} materials share
this floppiness, and it is related to the close-to-octahedral
coordination of the \ce{Ir} center and ensuing Jahn-Teller distortion.
From a practical standpoint, this causes the optimal geometrical
parameters of triplets to be sensitive to the chosen functional. To
minimize the impact of a functional, structures were optimized using
CAM-B3LYP~\cite{Yanai:2004/cpl/51} functional with the LANL2def-DZV
basis set~\cite{Hay:1985/jcp/299, Roy:2008/jctc/1029} using the
unrestricted formalism.
\begin{table}[t]
  \sisetup{ %
    table-format = 7.6, %
    round-mode = places, %
    round-precision = 6, %
    group-separator = {\,}, %
  } \centering
  \begin{tabularx}{1.0\linewidth}{@{}lSSSS}
    \toprule
    State & {Energy} & \multicolumn{3}{c}{Transition characteristics}
    \\ \cmidrule{3-5}
          &           & \multicolumn{1}{X}{\centering From orbital} & \multicolumn{1}{X}{\centering To orbital}  & \multicolumn{1}{X}{\centering Coefficient} \\
    \midrule
    1\ce{^3Ir(ppy)3}    & -1530.9249698867 & 126 & 131 & {0.2180} \\
    2\ce{^3Ir(ppy)3}    & -1530.9248621290 & 127 & 131 & {0.2166} \\
    3\ce{^3Ir(ppy)3}    & -1530.9246353622 & 130 & 131 & {0.2553} \\
    \bottomrule
  \end{tabularx}
  \caption{Energy (in \si{\hartree}) and the dominant orbital
    excitations for of the lowest states in the triplet manifold for
    \ce{Ir(ppy)3} in its S$_0$ configuration. The Fermi level is
    between orbitals 130 and 131, the reference energy is E(RHF) =
    \SI{-1531.0355568518}{\hartree}. }
  \label{tab:Ir_ppy3_triplet_manifold}
\end{table}

\section{Details of the Computational and Experimental Methods}
\label{sec:methods}

All \gls{DFT} calculations were run on a Westgrid high-performance
computing server. Each calculation was run on a single server node --
two-socket Intel E5-2683 v4 Broadwell@2.1GHz, 125~GB RAM, 960~GB SATA
SSD -- with allocated resources of 16 processor cores and a maximum of
64 GB RAM. Some calculations that required more memory (\emph{e.g.}
\gls{MP2}) were run with a maximum of 200 GB RAM. The actual memory
use varied depending on the compound under study and a type of
calculations performed. Values of T$_1$ → S$_0$ transition energy for
all methods employed are listed in Table~\ref{tab:energyvals}.

The \gls{iQCC} simulations were run on Ev3 Mircosoft Azure compute
instances that have Intel Xeon E5-2673 v4 Broadwell@2.3 GHz, 432 GB
RAM, and 864 GB SSD. The \gls{iQCC} method was implemented in Julia
language ver.~1.5~\cite{Julialang} and ran mostly serially, with
several non-critical parts parallelized with thread parallelization
available in Julia (\textsf{Threads.@parallel} with \textsf{nproc} =
16--32). All simulations finished between 200--480 CPU hours per
simulation. All simulations were run until the energy changes were
about $\SI{1e-5}{\hartree}$, which typically required $\sim 75$
iterations with 8 top-rank generators selected at each iteration
ranked by Eq.~\eqref{eq:EN1_amplitude_formula} (see
Sec.~\ref{sec:glsqcc-oper-rank}).

To assess the quantum circuit specifications of the QCC Ansatz,
ProjectQ was used to generate circuits and resource estimates
~\cite{Damian:2018/quantum/49}. Each of the nine Ir(III) complexes had
a total of 600 entanglers optimized and it was observed that most of
the high ranking entanglers were either 2-qubit or 4-qubit operators.
These entanglers were decomposed using the Time Evolution operator in
ProjectQ to generate a proposed quantum circuit. In this instance, the
longest circuit had a total of 3600 CNOT gates and 600 RZ rotation
gates. At the time of writing this value exceeds what is currently
possible on quantum hardware.

Empirical values for the T$_1$ → S$_0$ transitions for the nine Ir
(III) complexes were taken from the work of Sajoto \textit{et
  al.}~\cite{Sajoto:2009/jacs/9813}. As calibration, the
phosphorescence spectrum of \ce{Ir(ppy)_3} was confirmed with an
Edinburgh Instruments FS5 spectrofluorometer using the SC-70 sample
module. The spectrum was measured by dissolving \ce{Ir(ppy)_3} in
2-methyltetrahydrofuran at \SI{0.05}{\milli\gram\per\milli\litre}
mg/mL and cooling to \SI{77}{\kelvin} in a quartz EPR tube, then
exciting with \SI{365}{\nano\meter} light. The value obtained for the
T$_1$ → S$_0$ transition matched exactly
($\delta \lambda < \SI{1}{\nano\meter}$) with
Sajoto~\cite{Sajoto:2009/jacs/9813}.
\begin{landscape}
  \begin{table}
    \sisetup{ %
      table-format = 3.3, %
    } \centering %
    \scriptsize
    \begin{tabularx}{1.0\linewidth}{@{}lSSSSSSSSS}
      \toprule
      Method & \multicolumn{1}{c}{\ce{Ir(ppy)_3}}  & \multicolumn{1}{c}{\ce{Ir(F_2ppy)_3}} & \multicolumn{1}{c}{\ce{Ir(ppz)_2(ppy)}} & \multicolumn{1}{c}{\ce{Ir(F_2ppz)_2(ppy)}} & \multicolumn{1}{c}{\ce{Ir(ppz)_2(F_2ppy)}}  & \multicolumn{1}{c}{\ce{Ir(F_2ppz)_2(F_2ppy)}} & \multicolumn{1}{c}{\ce{Ir(ppz)_3}} & \multicolumn{1}{c}{\ce{Ir(F_2ppz)_3}} & \multicolumn{1}{c}{\ce{Ir(pmb)_3}} \\
      \midrule
      ROHF/RHF                          & 3.497 & 2.891 & 3.018 & 2.758 & 2.730 & 2.871 & 3.001 & 2.979 & 2.666\\
      UHF/RHF                            &  1.990 & 2.607 & 2.028 & 1.643 & 1.532 & 2.859 & 2.122 & 2.256 & 2.571\\
      RO-B3LYP/R-B3LYP              & {Failed} & {Failed} & {Failed} & 2.434 & {Failed} & 2.549 & {Failed} & 2.809 & 2.616\\
      U-B3LYP/R-B3LYP                & 2.630 & 2.454 & 2.484 & 2.366 & 2.247 & 2.471 & 2.622 & 2.734 & 2.560\\
      RO-PBE/RO-PBE                  & 3.993 & 4.465 & 5.375 & 2.264 & 5.298 & 2.369 & 4.250 & 4.752 & 3.975\\
      RO-CAMB3LYP/R-CAMB3LYP & 3.028 & 2.724 & 2.747 & 2.610 & 2.473 & 2.729 & 2.853 & 2.973 & 2.802\\
      U-CAMB3LYP/R-CAMB3LYP   & 3.133 & 2.583 & 2.629 & 2.483 & 2.331 & 2.582 & 2.730 & 2.841 & 2.715\\
      RO-LC-wPBE/R-LC-wPBE      & 3.119 & 2.833 & 2.760 & 2.704 & 2.496 & 2.828 & 2.916 & 3.063 & 2.902\\
      U-LC-wPBE/R-LC-wPBE        & 3.431 & 2.592 & 2.599 & 2.479 & 2.273 & 2.579 & 2.745 & 2.844 & 2.776\\
      TDDFT R-B3LYP                  & 2.502 & 2.294 & 2.288 & 2.213 & 1.980 & 2.302 & 2.400 & 2.525 & 2.393\\
      TDDFT + TDA R-B3LYP       & 2.549 & 2.430 & 2.381 & 2.334 & 2.046 & 2.450 & 2.492 & 2.681 & 2.470\\
      TDDFT R-CAMB3LYP           & 2.724 & 2.228 & 2.384 & 2.150 & 2.330 & 2.212 & 2.608 & 2.459 & 2.586\\
      TDDFT + TDA R-CAMB3LYP & 2.958 & 2.556 & 2.717 & 2.467 & 2.536 & 2.555 & 2.852 & 2.792 & 2.776\\
      TDDFT R-LC-wPBE              & 2.492 & 1.938 & 2.121 & 1.826 & 2.267 & 1.907 & 2.536 & 2.202 & 2.510\\
      TDDFT + TDA R-LC-wPBE   & 3.087 & 2.576 & 2.814 & 2.490 & 2.784 & 2.568 & 3.006 & 2.828 & 2.967\\
      ROMP2/RMP2                    & 8.901 & 2.926 & 3.611 & 2.768 & 2.317 & 2.911 & 3.017 & 3.495 & 3.369\\
      UMP2/MP2                        & 8.901 & 5.615 & 4.534 & 6.899 & 5.096 & 5.322 & 5.170 & 4.384 & 5.629\\
      \gls{iQCC}                        & 2.858 & 2.842 & 2.101 & 2.693 & 2.337 & 2.710 & 2.461 & 3.109 & 2.951\\
      \gls{iQCC} + PT                & 2.901 & 9.899 & 2.820 & 2.948 & 2.782 & 3.064 & 2.969 & 3.320 & 3.070\\
      \bottomrule
    \end{tabularx}
    \caption{T$_1$ → S$_0$ transitions of the Ir (III) complexes in eV
      dependent on computational method}
    \label{tab:energyvals}
  \end{table}
\end{landscape}

\bibliography{matsim-snap,encorr-loc,kspots-snap,qcomp-snap,books,programs,ekt-snap,databases}